\documentclass[aps,pre,twocolumn,showpacs,superscriptaddress]{revtex4}
\usepackage{amsfonts,amssymb,amsmath,graphicx,enumerate,color,times}

\begin{document}

\title{Identifying complex periodic windows in continuous-time dynamical systems using recurrence-based methods}

\author{Yong Zou}
    \affiliation{Potsdam Institute for Climate Impact Research, P.O. Box 601203, 14412 Potsdam, Germany}
\author{Reik V. Donner}
    \affiliation{Potsdam Institute for Climate Impact Research, P.O. Box 601203, 14412 Potsdam, Germany}
    \affiliation{Max Planck Institute for Physics of Complex Systems, N\"othnitzer Str.~38, 01187 Dresden, Germany}
    \affiliation{Institute for Transport and Economics, Dresden University of Technology, W\"urzburger Str.~35, 01187 Dresden, Germany}
\author{Jonathan F. Donges}
    \affiliation{Potsdam Institute for Climate Impact Research, P.O. Box 601203, 14412 Potsdam, Germany}
    \affiliation{Department of Physics, Humboldt University Berlin, Newtonstr.~15, 12489 Berlin, Germany}
\author{Norbert Marwan}
    \affiliation{Potsdam Institute for Climate Impact Research, P.O. Box 601203, 14412 Potsdam, Germany}
\author{J\"urgen Kurths}
  \affiliation{Potsdam Institute for Climate Impact Research, P.O. Box 601203, 14412 Potsdam, Germany}
    \affiliation{Department of Physics, Humboldt University Berlin, Newtonstr.~15, 12489 Berlin, Germany}

\date{\today}

\begin{abstract}
  The identification of complex periodic windows in the
  two-dimensional parameter space of certain dynamical systems has
  recently attracted considerable interest.  While for discrete
  systems, a discrimination between periodic and chaotic windows can
  be easily made based on the maximum Lyapunov exponent of the system,
  this remains a challenging task for continuous systems, especially
  if only short time series are available (e.g., in case of
  experimental data). In this work, we demonstrate that nonlinear
  measures based on recurrence plots obtained from such trajectories
  provide a practicable alternative for numerically detecting shrimps.
  Traditional diagonal line-based measures of recurrence
  quantification analysis (RQA) as well as measures from complex
  network theory are shown to allow an excellent classification of
  periodic and chaotic behavior in parameter space. Using the
  well-studied R\"ossler system as a benchmark example, we find that
  the average path length and the clustering coefficient of the
  resulting recurrence networks (RNs) are particularly powerful
  discriminatory statistics for the identification of complex periodic
  windows.

\end{abstract}

\maketitle

{\bf The investigation of the qualitative behavior in the full
  parameter space of a complex system is a very important, but often
  challenging task.  Detailed knowledge about the different possible
  types of dynamics helps understanding under which conditions
  particular states of a system lose stability or undergo significant
  qualitative changes.  In particular, in experimental settings, the
  availability of information about the underlying patterns in phase
  space allows tuning the critical parameters in such a way that one
  may obtain the desired working conditions.  Mathematically, the
  corresponding problem is traditionally investigated by means of
  bifurcation theory, which allows studying the properties of
  dynamical transitions in some
  detail~\cite{BifurTheory,Kuznetsov2004}.  However, the applicability
  of available methods for identifying bifurcation scenarios and
  determining the parameters at which they take place does often
  depend on the considered system itself. This is especially the case
  when dealing with larger sets of parameters, i.e., operating in a
  two- or even higher-dimensional parameter space, in particular for
  the case of experimental data.  In this work, we propose some
  methods based on recurrence properties in phase space that allow
  quantifying dynamically relevant properties from available time
  series, which we harness to disentangle the labyrinthine parameter
  space with respect to qualitatively and quantitatively different
  dynamics.  }

\section{Introduction}
The parameter space of nonlinear dynamical systems often exhibits a
rich variety of qualitatively different types of behavior, such as
periodic and chaotic regimes. Especially if more than one parameter is
responsible for the resulting complex bifurcation scenario, settings
leading to the same types of dynamics are often mutually entangled in
a rather complicated way. A specific example are so-called {\em
  shrimps}, a specific kind of periodic windows embedded in chaotic
regimes in the two-dimensional parameter space of a large class of
systems, which are characterized by a distinct structure consisting of
a main body and four thin legs~\cite{Gallas1,Gallas2}. The exact
properties of such structures however depend on the specific system
under study. In general, shrimps often display self-similarity and are
regularly organized along some distinguished
directions~\cite{Bonatto_prl_2005}. The particular orientation depends
on the respective stability conditions. When crossing the borders of
shrimps in different directions, the system typically shows different
bifurcation scenarios from periodic dynamics to chaos, e.g., via
period doubling or via intermittency~\cite{Gallas2,Zou_ijbc_2005}. A
detailed stability analysis of the periodic solutions contained within
shrimps, therefore, presents a promising approach for understanding
the dynamical backbone of these structures. Consequently, the study of
this particular type of structure has recently attracted considerable
interest.

The emergence of shrimps has firstly been described in great detail
for chaotic
maps~\cite{Gallas1,Gallas2,EBarreto,Lorenz_physD_2008,Martins2008},
although corresponding structures have already been reported in
earlier studies for both maps and time-continuous
systems~\cite{Perez1982,Belair1983,Gaspard_jsp1984,Belair1985}.  In
the last years, additional efforts have been spent on numerically
identifying such structures in systems of ordinary differential
equations (ODEs) as well~\cite{MBthesis}. Examples include the
R\"ossler
system~\cite{Gaspard_jsp1984,MTthesis,Bonatto2008PTRS,Gallas_ijbc_2010},
integrate-and-fire models of neurophysiological
oscillations~\cite{Glass2001}, two coupled parametrically excited
oscillators~\cite{Zou_ijbc_2005}, models of laser
dynamics~\cite{Bonatto_prl_2005,Bonatto_pre_2007,bonatto_prl_2008,Bonatto2008PTRS,Kovanis_epjd_2010},
the damped-driven Duffing system~\cite{bonatto_pre_2008}, different
nonlinear electronic
circuits~\cite{Bonatto2008PTRS,Albuquerque2008,Albuquerque2009,Cardoso2009,Viana_chaos_2010},
chemical
oscillators~\cite{Bonatto2008PTRS,Gallas_ijbc_2010,Freire2009}, the
Lorenz-84 low-order atmospheric circulation
model~\cite{Bonatto2008PTRS}, or a vibro-impact
system~\cite{deSouza2009}. Recently, shrimp structures have been
experimentally found in the chaotic
Chua~\cite{Baptista2003,maranh_pre_2008} and Nishio-Inaba
circuits~\cite{Stoop_prl_2010}, which underlines the increasing
importance of proper numerical algorithms for automatically
identifying such periodic windows in both theoretical and experimental
studies, possibly even in the presence of noise.

Typically, at the boundaries of a shrimp, small inaccuracies of the
parameters induce dramatic changes in the resulting dynamics.
Therefore, these structures can hardly be uncovered by existing
analytical methods based on linear stability analysis, impossible in
particular at the boundaries. In most recent works for continuous
systems, the maximum Lyapunov exponent $\lambda_{1}$ has been
estimated numerically, which allows distinguishing periodic and
chaotic dynamics since $\lambda_1=0$ for periodic orbits and
$\lambda_1>0$ for chaotic ones. Recently, an alternative method has
been proposed for uncovering shrimps in systems described by
ODEs~\cite{MTthesis} based on the concept of correlation entropy
$K_{2}$, which can be considered as a lower bound for the sum of all
positive Lyapunov exponents of the system~\cite{Kantz97}. It has been
demonstrated that considering $K_2$ indeed yields results of
comparable accuracy as the maximum Lyapunov exponent
method~\cite{Zou_ijbc_2005} and is more efficient, because less data
points are needed.

A convenient way for numerically estimating $K_{2}$ is using {\em
  recurrence plots} (RPs), an efficient technique of nonlinear time
series analysis. Given a trajectory of a dynamical system consisting
of different values $\mathbf{x}_i$, where $i$ indicates the time of
observation, the corresponding RP is defined
as~\cite{Eckmann,Marwan_report_2007}
\begin{equation} \label{rp_definition}
R_{i,j}(\varepsilon)=\Theta(\varepsilon-\|\mathbf{x}_i-\mathbf{x}_j \|),
\end{equation}
where $\Theta(\cdot)$ is the Heaviside function,
$\|\mathbf{x}_i-\mathbf{x}_j \|\equiv d_{i,j}$ is the distance between
two observations $\mathbf{x}_i$ and $\mathbf{x}_j$ in phase space
(which will be measured in terms of the maximum norm in the
following), and $\varepsilon$ a pre-defined threshold for the
proximity of two states in phase space, i.e., for distinguishing
whether or not two observations are neighbors in phase space. Hence,
the basic mathematical structure describing a RP is the binary
recurrence matrix $R_{i,j}$. Visualizing this matrix by black
($R_{i,j}=1$) and white ($R_{i,j}=0$) dots, different types of
dynamics can be identified in terms of different features of line
structures (including discrete points, blocks, and extended diagonal
or vertical lines), which can be quantitatively assessed in terms of
recurrence quantification analysis (RQA, see Sec.~\ref{sec:rqa}).

The recurrence plot based estimation of $K_{2}$ involves two main
steps: (i) computing the cumulative probability distributions of the
lengths of diagonal lines, and (ii) properly selecting a scaling
region in dependence on the diagonal line length $l$ and evaluating
its characteristic parameters~\cite{fluid}. Since the second step
assumes the existence of sufficient convergence in a reasonable
regime, in practical applications various values of $\varepsilon$ need
to be considered for properly detecting the corresponding scaling
region. Hence, this approach is partially subjective and depends on
the specific choice of the scaling region.

In this work, we suggest and thoroughly study alternative approaches
to analyze quantitatively the corresponding properties of the
recurrence matrix.  More specifically, we apply measures from complex
network theory to the recurrence structures and compare their
performance with that of more traditional RQA measures. The
corresponding framework of recurrence networks (RNs), i.e., the idea
of interpreting the recurrence matrix $R_{i,j}$ of
Eq.~(\ref{rp_definition}) as the adjacency matrix $A_{i,j}$ of an
undirected, unweighted complex network by setting
\begin{equation}
  A_{i,j}=R_{i,j} - \delta_{i,j},
\end{equation}
where $\delta_{i,j}$ is the Kronecker delta, has been recently
proposed for identifying dynamical transitions in model systems such
as the logistic map~\cite{Marwan2009}, as well as detecting subtle
changes in a marine dust flow record.  It should be noted that similar
approaches of understanding the properties of time series from complex
network perspectives have been suggested in parallel by various
authors~\cite{Zhang2006,Xu2008} (see \cite{Donner2009,Donner2011} for
further references).

While most classical RQA measures are based on line structures in the
RP, complex network measures reveal higher-order properties of the
phase space density of states of the considered dynamical
system~\cite{Donner2009} (see Sec.~\ref{sec:net}). However, the
question whether the complex network approach is more suitable than
conventional RQA for distinguishing different types of dynamics has
not yet been systematically addressed. In this work, we provide a
corresponding comparative analysis using some particularly well-suited
measures of both types. Specifically, we demonstrate clear advantages
of complex network measures in identifying shrimp structures or, even
more, arbitrary complex periodic windows in the two-dimensional
parameter space of certain dynamical systems by taking the
well-studied R\"ossler system as an illustrative benchmark example.
The performance of different measures is compared by means of several
statistical tests based on the resulting patterns in the parameter
space.

The remainder of this paper is organized as follows: In
Sec.~\ref{sec:method}, we briefly review some RQA as well as complex
network measures and outline important technical issues arising when
using these measures, in particular, the dependence on the choice of
the recurrence threshold $\varepsilon$ and other parameters. A
specific part of the two-dimensional parameter space of the R\"ossler
system that is known to contain shrimps is examined in
Sec.~\ref{sec:lyap} by means of the maximum Lyapunov exponent
$\lambda_{1}$, which serves as a reference for our results obtained
using recurrence-based methods in Sec.~\ref{shrimp_num}. The main
conclusions of our work are summarized in Sec.~\ref{sec:conclusion}.

\section{Methods} \label{sec:method}

\subsection{Recurrence quantification analysis (RQA)}\label{sec:rqa}
Recurrence quantification analysis (RQA) has been introduced for
quantifying the presence of specific structures in RPs
\cite{Zbilut1992,Trulla1996,Marwan2008,Webber2009}. During the last
years, numerous methodological developments and applications in
various fields of science have been reported (for a corresponding
review, see~\cite{Marwan_report_2007}). Most traditional RQA measures
are based on the length distributions of diagonal or vertical lines.
In this work, we will particularly make use of the following three
quantities:
\begin{itemize}
\item The \textbf{recurrence rate}
\begin{equation}
RR=\frac{2}{N(N-1)} \sum_{i<j} R_{i,j}, \label{eq:rr}
\end{equation}
measures the fraction of recurrence points in a RP and, hence, gives
the mean probability of recurrences in the system.

\item The \textbf{determinism} $DET$ is defined as the percentage of
  recurrence points belonging to diagonal lines of at least length
  $l_{min}$ (see Sec.~\ref{sec:lmin} for details),
\begin{equation} \label{det_hamiltonian}
DET = \frac{\sum_{l=l_{min}}^N l P(l)}{\sum_{l=1}^N l P(l)},
\end{equation}
where $P(l)$ denotes the probability of finding a diagonal line of
length $l$ in the RP. This measure quantifies the predictability of a
system.

\item The \textbf{average diagonal line length} $L_{MEAN}$, defined as
\begin{equation} \label{lmean_hamiltonian}
L_{MEAN} = \frac{\sum_{l=l_{min}}^N l P(l) }{\sum_{l=l_{min}}^N P(l) },
\end{equation}
characterizes the average time that two segments of a trajectory stay
in the vicinity of each other, and is related to the mean
predictability time.
\end{itemize}

In addition to (diagonal as well as vertical) line-based RQA measures,
in some cases, parameters characterizing recurrence times (i.e.,
vertical distances between two recurrence points) based on the RPs
have been suggested to be suitable for quantifying basic properties of
the recorded
dynamics~\cite{Zou_quasiperiod,jojo2007pre,jojo2008chaos}.

\begin{figure}[thb]
  \centering
  \includegraphics[width=0.45\textwidth]{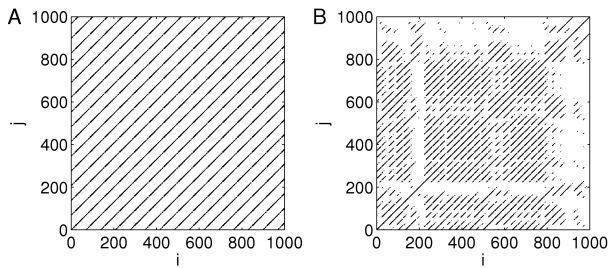}
  \caption{\small {Recurrence plots for a periodic (A) and a chaotic
      (B) trajectory of the R\"ossler system (\ref{ros_eqs}) (see
      Sec.~\ref{two_orbits} for details). } } \label{rp_pc}
\end{figure}

Let us now consider in what sense RQA measures behave differently in
the presence of periodic and chaotic dynamics. In a RP, a periodic
orbit is reflected by long non-interrupted diagonal lines that are
separated by a constant offset, which corresponds to the period of the
oscillation (Fig.~\ref{rp_pc}A). In contrast, for a chaotic
trajectory, the distances between diagonal lines are not constant due
to multiple time scales present in the system (Fig.~\ref{rp_pc}B).
Furthermore, the existing diagonals are interrupted because of the
exponential divergence of nearby trajectories. Therefore, we expect
that both $DET$ and $L_{MEAN}$ typically have larger values for a
periodic trajectory than for a chaotic one.

\subsection{Quantitative analysis of recurrence networks}\label{sec:net} 
Recurrence networks (RNs) are spatial networks representing local
neighborhood relationships in the phase space of a dynamical system.
In a RN, every sampled point is represented by a vertex $v$, whereas
edges indicate pairs of observations whose mutual distance in phase
space is smaller than the pre-defined threshold $\varepsilon$. The
resulting edge density $\rho$, i.e., the relative fraction of edges
that are actually present in comparison to a fully connected network,
is then given by $RR$ (see Eq.~(\ref{eq:rr})).  Hence, characterizing
topological properties of RPs by sophisticated network-theoretic
quantifiers allows retrieving additional information about
higher-order phase space properties of the system~\cite{Donner2009}.
It has been shown that these measures are able to capture dynamical
transitions in complex systems, such as in the logistic map with
changing control parameter or a real-world paleoclimatic time series,
demonstrating that network-theoretic features provide additional and
complementary information when compared to traditional RQA
measures~\cite{Marwan2009}.

In this work, we particularly consider the global clustering
coefficient $\mathcal{C}$ and the average path length $\mathcal{L}$ of
a RN. The application of other measures (e.g., betweenness centrality)
is straightforward, but we argue that the description of network
topology by a scalar parameter instead of a distribution of vertex- or
edge-based statistics might be more suitable for detecting and
quantifying qualitative changes in the dynamics of a system. Moreover,
one has to note that state-of-the-art numerical algorithms for
computing other complex network measures often require considerably
larger computational efforts than estimating traditional RQA measures.

The local clustering coefficient $\mathcal{C}_v$ characterizes the
phase space geometry of an attractor in the $\varepsilon$-neighborhood
of a vertex $v$. Specifically, $\mathcal{C}_v$ gives the probability
that two randomly chosen neighbors of $v$ are also
neighbors~\cite{watts_nature_1998}, i.e.,
\begin{equation}
\mathcal{C}_v=\frac{2}{k_v(k_v-1)} N^{\Delta}_v, \label{cc_local}
\end{equation}
where $k_v$ is the degree centrality (i.e., the number of
neighbors of $v$, which coincides with the local recurrence rate
$RR_v$), and $N^{\Delta}_v$ is the total number of closed
triangular subgraphs including $v$, which is normalized by the
maximum possible value $k_v(k_v-1)/2$. For vertices of degree
$k_v=0$ or $1$ (i.e., isolated or tree-like vertices,
respectively), $\mathcal{C}_v=0$ by definition since such vertices
cannot participate in triangles. Instead of studying
$\mathcal{C}_v$ individually for all vertices of a RN, we consider
its average value taken over all vertices of a network, the
\textbf{global clustering coefficient}
\begin{equation}
\mathcal{C}=\frac{1}{N} \sum_{v=1}^N \mathcal{C}_v,
\end{equation}
as a global characteristic parameter of network topology.

The local clustering coefficient $\mathcal{C}_v$ is related to the
effective dimensionality of the set of observations in the
$\varepsilon$-neighborhood of a vertex $v$~\cite{Donner2009}. In
particular, we find specific dependencies for both continuous and
discrete dynamical systems:
\begin{enumerate}[(i)]
\item For discrete systems, periodic orbits consist only of a finite
  set of points. This implies that for $\varepsilon \ll \Delta$ with
  $\Delta \ge \max_{i,j}d_{i,j}$ being the attractor diameter, the RN
  is decomposed into disjoint components with multiple vertices being
  located at the same point in phase space.  As a consequence, the
  individual components are fully connected for all $\varepsilon>0$,
  which leads to $\mathcal{C}=1$. In contrast, for chaotic
  trajectories, sufficiently small $\varepsilon\ll \Delta$ results in
  $\mathcal{C}<1$~\cite{Donner2009}.

\item For a continuous system, it is a well-established fact that if a
  chaotic trajectory enters the neighborhood of an unstable periodic
  orbit (UPO), it stays within this neighborhood for a certain
  time~\cite{Lathrop_pra_1989}. As a consequence, states accumulate
  along this UPO instead of homogeneously filling the phase space in
  the corresponding neighborhood (in particular if we consider UPOs of
  lower period).  Since from the theory of spatial random
  graphs~\cite{Dall2002} it is known that the average clustering
  coefficient increases with decreasing spatial dimension of the space
  in which a network is embedded, it follows that parts of a chaotic
  attractor that are close to (low-periodic) UPOs are characterized by
  higher values of $\mathcal{C}_v$ in the resulting
  RNs~\cite{Donner2009}. Following a related argument, we expect that
  states on a periodic orbit of a continuous system are also
  characterized by high values of $\mathcal{C}_v$, which implies high
  $\mathcal{C}$. In general, however, for chaotic trajectories the
  filling of the phase space with observed states is more homogeneous
  than for periodic ones, which leads to a tendency towards lower
  values of $\mathcal{C}_v$ and, hence, $\mathcal{C}$.
\end{enumerate}

\begin{figure}[thb]
  \centering
  \includegraphics[width=0.45\textwidth]{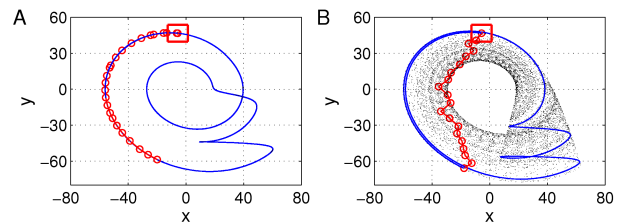}
  \caption{\small{(Color online) Illustration of the shortest path
      length for two example trajectories of the R\"ossler system
      (\ref{ros_eqs}) in (A) periodic and (B) chaotic regime. The
      square is a schematic projection of the recurrence neighborhood
      to the $(x,y)$ plane. In these two particular examples,
      $l^A_{i,j} = 24$ and $l^B_{i,j} = 19$ for $\varepsilon = 7.0$
      (maximum norm). } \label{path_pc} }
\end{figure}

A second global measure of network topology is the \textbf{average
  path length}. Since we consider RNs as undirected and unweighted, we
define all edges to be of unit length in terms of graph (geodesic)
distance. The graph distance between any two vertices of the network
is defined as the length of the shortest path between them. Hence, the
shortest path length $l_{i,j}$ in the RN gives the minimum number of
edges that have to be passed on a graph between two vertices $i$ and
$j$. Accordingly, $l_{i,j}$ is related to the phase space distance of
the associated states, but \textit{not} to the temporal evolution of
the system (Fig.  \ref{path_pc}).

The average path length $\mathcal{L}$ is defined as the mean value of
the shortest path lengths $l_{i,j}$ taken over all pairs of vertices
$(i,j)$,
\begin{equation}
\mathcal{L} = \left<l_{i,j}\right>=\frac{2}{N(N-1)}\sum_{i<j} l_{i,j}.
\end{equation}
For disconnected pairs of vertices, the shortest path length is set to
zero by definition. Note that in most practical applications, this has
no major impact on the corresponding statistics. The average path
length is related to the average separation of states in phase space,
which measures the size of the attractor in units of $\varepsilon$.
Note that since metric distances in the phase space are conserved,
this implies an approximately reciprocal relationship
$\mathcal{L}\propto\varepsilon^{-1}$~\cite{Donner2009}, with the
exception of periodic orbits of discrete maps (see the arguments given
above).

For the behavior of the average path length $\mathcal{L}$, one has to
carefully distinguish between continuous and discrete systems again:
\begin{enumerate}[(i)]
\item For discrete systems, the structure of a periodic orbit in phase
  space implies $\mathcal{L}=1$ by definition for $\varepsilon \ll
  \Delta$~\cite{Marwan2009}. In turn, for chaotic trajectories, there
  is a continuum of possible values, so that $\mathcal{L}>1$ for
  $\varepsilon \ll \Delta$. Hence, the average path length of a
  periodic orbit is smaller than that of a chaotic one.
\item For continuous systems, however, a periodic trajectory has a
  completely different structure in phase space, which means that
  $\mathcal{L}$ is approximately determined by the curve length of the
  orbit in the phase space (Fig.~\ref{path_pc}A). In contrast, for
  chaotic trajectories, there are ``short-cuts'' that allow reaching
  one position in phase space from another with a lower number of
  steps than by following the trajectory~\cite{Donner2009}
  (Fig.~\ref{path_pc}B). Consequently, for a comparable attractor
  diameter $\Delta$ and the same recurrence threshold $\varepsilon$,
  $\mathcal{L}$ can be expected to have lower values for chaotic
  orbits. Alternatively, it is desirable to fix $RR$ instead of
  $\varepsilon$ to obtain RNs with approximately the same numbers of
  edges. A detailed discussion on the advantages of this approach is
  presented in Sec.~\ref{sec:rr}. In this case, we note that the
  different geometric structure of the attractor in both the periodic
  and chaotic regime even enhances the effect of shortcuts, so that
  under general conditions, periodic orbits of continuous systems are
  characterized by larger $\mathcal{L}$ than chaotic ones.
\end{enumerate}

Concerning the behavior of the average path length as shown in
Fig.~\ref{path_pc}, we have to point out that differences between
periodic and chaotic dynamics require that transient behavior (e.g.,
during some initial phase of the dynamics before the attractor has
been reached by the considered trajectory) has been sufficiently
removed, since such transients could also result in artificial
short-cuts in phase space for periodic systems. While this problem can
be easily solved in numerical studies by simply discarding the
transients, it can contribute additional uncertainties in experimental
studies where only a limited amount of data is available. In a similar
way, using $\mathcal{L}$ for discriminating between periodic dynamics
and chaos in noisy experimental time series is expected to lead to the
same problem. However, in such cases, other network-theoretic
quantities (e.g., $\mathcal{C}$) still provide feasible alternatives,
to which the mentioned conceptual problem does not apply (at least for
sufficiently small noise levels that do not exceed the order of
magnitude of the recurrence threshold $\varepsilon$). In addition,
there are other recurrence-based techniques such as recurrence time
statistics, which allow distinguishing regular and chaotic dynamics
even in challenging situations, e.g., quasi-periodicity versus weakly
chaotic behavior~\cite{Zou_quasiperiod,Zou2007Chaos}.

In summary, $\mathcal{C}$ shows a consistent behavior for discrete and
continuous systems, whereas $\mathcal{L}$ performs inversely with
respect to periodic and chaotic dynamics. This implies, that
$\mathcal{L}$ cannot be used alone to \textit{classify} the dynamics
of time series from real systems, where the nature of the underlying
process is unknown. Our results presented in this work and
elsewhere~\cite{Marwan2009} suggest, however, that $\mathcal{L}$ is
nevertheless very well able to \textit{distinguish} between periodic
and chaotic behavior in the parameter space of the same dynamical
system.

\section{Model: Chaotic R\"ossler system} \label{sec:lyap}

\subsection{Two-dimensional parameter space}
As an illustrative example for a time-continuous dynamical system that
shows shrimp structures in its parameter space, we study the R\"ossler
system
\begin{equation} \label{ros_eqs}
\left(\frac{dx}{dt},\frac{dy}{dt},\frac{dz}{dt}\right) =
\left(-y-z, x+ay, b+z(x - c)\right).
\end{equation}
In this system, $c$ is often regarded as the standard bifurcation
parameter, whereas $a$ and $b$ mainly change the shape of the
attractor. In this work, however, we vary $a=b$ on the one hand, and
$c$ on the other hand, which yields a two-dimensional parameter space
of $(c,a)$. One of the parameter regions of special interest is $(c,a)
\in [20,45] \times [0.2,0.3]$, where the chaotic regime is
intermingled with periodic windows in a complex
manner~\cite{MTthesis}. Thus, in the following, we pay special
attention to this part of the parameter space. Note that shrimp
structures are not confined to this particular plane in the
three-dimensional parameter space of the R\"ossler system and, hence,
investigating a differently oriented plane as
in~\cite{Bonatto2008PTRS} would not qualitatively alter the results of
our study.

For further analysis, the above mentioned part of the parameter space
of $(c,a)$ is divided into $1,000 \times 1,000$ grid points, which
results in the step size $0.0001$ in $a \in [0.2, 0.3]$ and $0.025$ in
$c \in [20, 45]$. We first consider the maximum Lyapunov exponent of
the resulting systems as a characteristic measure to determine whether
the trajectories resulting from each parameter combination are chaotic
or periodic. For this purpose, all three Lyapunov exponents
$\lambda_{i}$ have been computed based on Eq.~(\ref{ros_eqs}) by
numerically solving the associated variational
equations~\cite{Ott2002}. Numerical integration is carried out using a
fourth-order Runge-Kutta integrator with a fixed step size of
$h=0.001$ time units and randomly chosen initial conditions
$(x_0,y_0,z_0)\in [0,1]\times[0,1]\times[0,1]$.  In order to provide
sufficient data for an accurate estimation of Lyapunov exponents, the
integration is performed for sufficiently long time (until $t=10,000$,
i.e., involving $N=10,000,000$ data points with a sampling $\Delta t$
corresponding to the considered integration step $h$). For computing
the different recurrence-based measures in a second step of analysis,
the initial transients (cf.~Sec.~\ref{sec:net}) have been discarded by
removing the first $500,000$ iterations (i.e., all data until $t=500$)
from the simulated trajectories. Moreover, for all further
calculations based on the recurrence matrices, a coarser sampling of
$\Delta t=0.2$ (i.e., 200 integration steps) has been considered. For
each trajectory, the sampling has been terminated after $N=5,000$
points (corresponding to about $150 \sim 250$ oscillations of the
system) have been obtained, yielding much shorter time series than
those used for estimating Lyapunov exponents.  Note that the specific
choice of sampling time has a certain influence on the results
discussed in the following. Specifically, the sampling corresponding
to the optimum resolution of dynamically relevant features is expected
to vary among different parameter settings. In this respect, the
considered value of $\Delta t$ represents a reasonable and practically
tractable choice.
\begin{figure}[thb]
  \centering
  \includegraphics[width=0.45\textwidth]{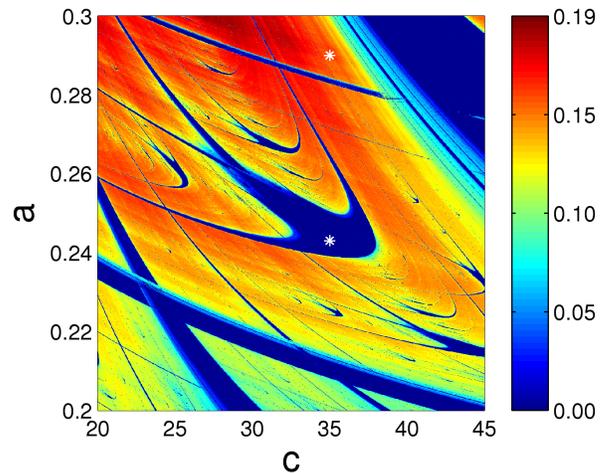}
  \caption{\small {(Color online) Maximum Lyapunov exponent
      $\lambda_1$ in the $(c, a)$ parameter plane of the R\"ossler
      system (\ref{ros_eqs}). Regions with $\lambda_{1} = 0$ indicate
      periodic dynamics, those with large $\lambda_1$ correspond to a
      strongly chaotic behavior. Asterisks indicate the parameter
      combinations used as examples in Sec.~\ref{sec:method}
      and~\ref{sec:lyap}.}
\label{lyap_ac} }
\end{figure}

The resulting maximum Lyapunov exponent $\lambda_{1}$ reveals a rich
structure of chaotic and periodic dynamics in the parameter space
(Fig.~\ref{lyap_ac}). Inside the chaotic regions, several well
pronounced periodic bands are identified. Moreover, in the center of
the plot, a special periodic window of interest is found, with
structures resembling a head and four main thin legs.  These
particular swallow-like structures can be identified as
shrimps~\cite{MTthesis}. Additional zooms into the parameter space
uncover numerous fine structures, which correspond to secondary
shrimps.

\subsection{Prototypical trajectories} \label{two_orbits}
\begin{figure}[thb]
  \centering
  \includegraphics[width=0.45\textwidth]{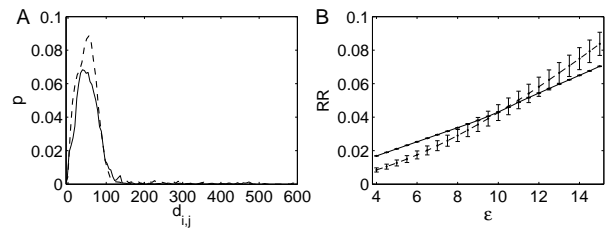}
  \caption{\small {(A) Probability distributions $p(d_{i,j})$ of
      mutual distances $d_{i,j}$ (maximum norm) between states on one
      realization ($N=5,000$) of periodic (solid) and chaotic (dashed)
      trajectories, respectively (see text). (B) Dependence of the
      recurrence rate $RR$ on the recurrence threshold $\varepsilon$
      for a periodic and chaotic regime. The error bars correspond to
      the standard deviation obtained from $100$ realizations with
      different initial conditions.} \label{rreps} }
\end{figure}

We first illustrate the potentials of recurrence-based approaches for
discriminating between periodic and chaotic dynamics by studying
trajectories obtained from two distinct parameter combinations. Based
on the parameter space shown in Fig.~\ref{lyap_ac}, the dynamics is
periodic for $(a,b,c)=(0.245,0.245,35)$, while chaotic for
$(a,b,c)=(0.29,0.29,35)$. The Lyapunov exponents are $(0, -0.223,
-32.683)$ for the periodic trajectory, and $(0.151,0,-32.517)$ for the
chaotic one. The typical phase space distances of states on the two
trajectories are of the same order of magnitude
(Tab.~\ref{tab_pechaotic}), which is illustrated by the probability
distribution of the mutual distances $d_{i,j}$ in Fig.~\ref{rreps}A.
However, the detailed relationship of $RR(\varepsilon)$ is clearly
different: for a low (but fixed) $RR$, we have to consider much higher
$\varepsilon$ for chaotic trajectories than for periodic ones, while
for larger $RR(\varepsilon)$, the opposite behavior is found
(Fig.~\ref{rreps}B). $RR$ corresponds to the correlation sum, the
slope of which in a double logarithmic plot gives the correlation
dimension $D_2$. Therefore, $RR(\varepsilon)$ having a larger slope
for the chaotic trajectory than for the periodic one
(Fig.~\ref{rreps}B) indicates that the correlation dimension of the
former is higher than that of the latter.
\begin{table}[thb]
  \centering
  \begin{ruledtabular}
  \begin{tabular}{c|c|c}
   & periodic & chaotic \\
  \hline
  $\lambda_1$ & $0.0\pm0.0003$ & $0.15\pm0.0002$ \\
  \hline
  $\left<d_{i,j}\right>$ & $58.95\pm0.14$ & $57.42\pm3.03$ \\
  $\max_{i,j} d_{i,j}$ & $476.69\pm0.15$ & $575.12\pm23.99$ \\
  $\varepsilon\left(RR=0.02\right)$ & $4.79\pm0.14$ & $6.50\pm0.53$ \\
  \hline
  $DET$ & $0.99\pm0.01$ & $0.97\pm0.01$ \\
  $L_{MEAN}$ & $18.51\pm0.18$ & $10.60\pm0.34$ \\
  \hline
  $\mathcal{C}$ & $0.77\pm0.001$ & $0.62\pm0.004$ \\
  $\mathcal{L}$ & $45.30\pm1.83$ & $9.19\pm0.37$ \\
  \end{tabular}
  \end{ruledtabular}
  \caption{\small {Maximum Lyapunov exponents $\lambda_1$
      ($N=10,000,000$, $\Delta t=0.001$), mean and maximum separation of
      points in phase space ($N=5,000$, $\Delta t=0.2$) and resulting
      recurrence threshold $\varepsilon$ (maximum norm) for $RR=0.02$,
      and RQA ($l_{min}=2$) and network measures for two parameter
      combinations (see text), representing periodic and chaotic
      regimes of the R\"ossler system. The error bars correspond to the
      standard deviation obtained from $100$ realizations with different
      initial conditions. Note that the large variance of the metric
      quantities $\varepsilon$ and $d_{ij}$ for the chaotic trajectory
      is a common result when working with short time series and
      different initial conditions~\cite{Donner2009b}.} }
\label{tab_pechaotic}
\end{table}

Both RQA and complex network measures highlight differences in the
topological structure between the periodic and chaotic regimes
(Tab.~\ref{tab_pechaotic}). We observe that for the considered
recurrence rate ($RR=0.02$), $DET$, $L_{MEAN}$, $\mathcal{C}$ and
$\mathcal{L}$ have lower values for chaotic trajectories than for
periodic ones (which distinctively differs from the behavior of the
maximum Lyapunov exponent $\lambda_1$), while the difference is
significant for the latter three measures. However, before being able
to generalize these findings, the robustness of the aforementioned
features has to be critically examined. Our corresponding results are
summarized in the following two subsections.

\subsection{Dependence on $l_{min}$} \label{sec:lmin}
\begin{figure}[thb]
  \centering
  \includegraphics[width=0.45\textwidth]{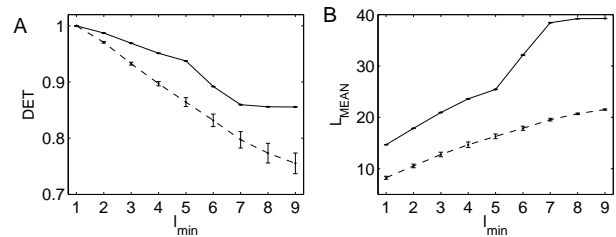}
  \caption{\small {Dependence of the RQA measures $DET$ (A) and
      $L_{MEAN}$ (B) on $l_{min}$ for periodic (solid) and chaotic
      (dashed) trajectories. The error bars indicate the standard
      deviation obtained from $100$ realizations of the R\"ossler
      system (\ref{ros_eqs}) with $N=5,000$, $\Delta t=0.2$,
      $RR=0.02$, and different initial conditions.} \label{lmin_rqa} }
\end{figure}

For the computation of most line-based RQA measures (with some
exceptions, such as $RR$ or the maximum diagonal line length), one has
to choose a proper minimal line length $l_{min}$, to avoid a bias due
to oversampling~\cite{Marwan_report_2007}. The particular values of
these measures do significantly depend on $l_{min}$
(Fig.~\ref{lmin_rqa}), although the discrepancy between the values of
$DET$ and $L_{MEAN}$ obtained for periodic and chaotic trajectories
remains qualitatively unchanged (in general, the values of $DET$ and
$L_{MEAN}$ are larger for a periodic trajectory than for a chaotic
one). Note that $DET$ and $L_{MEAN}$ are more robust against noise
than other measures like the maximum diagonal line
length~\cite{Marwan_report_2007}.

\subsection{Dependence on the recurrence threshold}\label{sec:rr}
Since all considered measures are based on the recurrence matrix,
their values necessarily depend on the choice of the recurrence
threshold $\varepsilon$. So far, there is no universal threshold
selection criterion for the RP computation. On the one hand, if
$\varepsilon$ is chosen too small, there are almost no recurrence
points and, hence, no feasible information on the recurrence structure
of the system. On the other hand, if $\varepsilon$ is too large,
almost every point is a neighbor of every other point, which leads to
numerous artifacts. Hence, we have to seek a compromise for choosing a
reasonable value of $\varepsilon$. One rule of thumb is choosing
$\varepsilon$ in such a way that it lies in the scaling regime of the
double logarithmic plot of the correlation integral versus
$\varepsilon$. This rule coincides with the classical strategy for
estimating the correlation dimension $D_2$ using the
Grassberger-Procaccia algorithm~\cite{Grassberger_prl_1983}. Following
independent arguments, Schinkel \textit{et~al.}~\cite{Schinkel2008}
suggested choosing $\varepsilon$ corresponding to at most $5\%$ of the
maximum attractor diameter in phase space. In a similar way,
$RR\lesssim 0.05$ has been found a reasonable choice in the analysis
of RNs \cite{Marwan2009,Donner2009,Donner2009b}.

\begin{figure}[thb]
  \centering
  \includegraphics[width=0.45\textwidth]{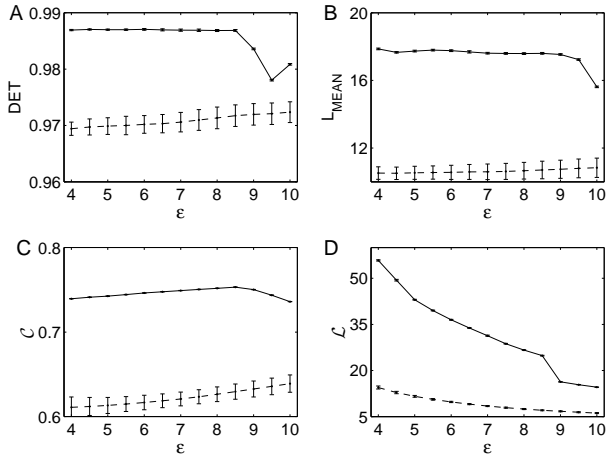}
  \caption{\small {Dependence of the RQA measures $DET$ (A),
      $L_{MEAN}$ (B), and the network measures $\mathcal{C}$ (C) and
      $\mathcal{L}$ (D) on the recurrence threshold $\varepsilon$ for
      periodic (solid) and chaotic (dashed) trajectories (see text).
      The error bars indicate the standard deviation obtained from
      $100$ realizations of the R\"ossler system (\ref{ros_eqs}) with
      $N=5,000$, $\Delta t=0.2$, $l_{min}=2$, and different initial
      conditions.}} \label{eps_rqa}
\end{figure}

The dependence of RQA and RN measures on $\varepsilon$ is shown in
Fig.~\ref{eps_rqa}. One observes that all previously discussed
measures are clearly able to distinguish between periodic and chaotic
dynamics, i.e., they show significant differences in a broad interval
of $\varepsilon$ (corresponding to mean recurrence rates of about 1\%
to 5\%).

\begin{figure}[thb]
  \centering
  \includegraphics[width=0.45\textwidth]{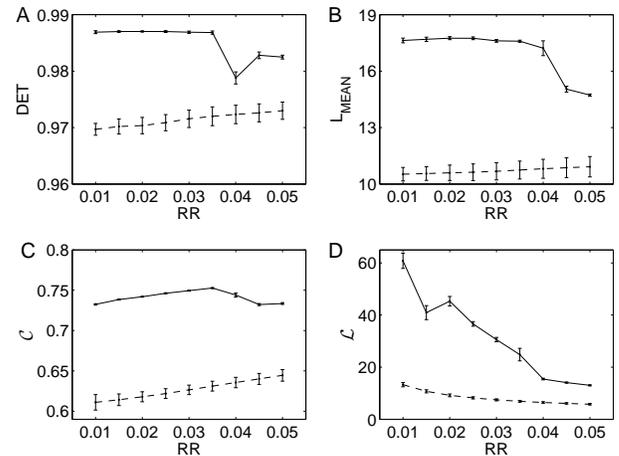}
  \caption{\small {Same as in Fig.~\ref{eps_rqa} for the dependence of
      these measures on the recurrence rate $RR$. }} \label{rr_rqa}
\end{figure}

Instead of fixing the recurrence threshold $\varepsilon$, it may be
desirable to compare different situations by using RPs with a fixed
value of $RR$. Firstly, the resulting RNs have approximately the same
number of edges, which allows comparing the resulting topological
properties of different networks more objectively.  Secondly, it has
been shown in Fig.~\ref{rreps}B that there is a crossover between the
$RR$ of periodic and chaotic trajectories obtained with the same
$\varepsilon$, which is related to the fact that $RR(\varepsilon)$
(which corresponds to the correlation sum) increases exponentially for
a chaotic trajectory~\cite{Grassberger_prl_1983}. Finally, we have to
note that the amplitudes of the (chaotic or periodic) oscillations of
the R\"ossler system change with the specifically chosen control
parameters of the system. In order to automatize the numerical
algorithms for estimating all RP-based measures, fixing $\varepsilon$
would lead to values of $RR$ and other RQA measures that cannot be
compared in a meaningful way. A related approach using the dependence
on $RR$ has already been applied for an automatized detection of the
scaling region in the recurrence-based estimation of $K_2$ for the
study of a large scale system~\cite{Planetary}.

According to the above arguments, we will use a fixed value of $RR$ in
all following calculations. Because of this, in a first step of
analysis, we study the dependence of the different measures on $RR$ in
some detail similar to the dependence on $\varepsilon$ discussed
above. Figure~\ref{rr_rqa} reveals that the difference between
periodic and chaotic orbits remains qualitatively unchanged and does
not depend on the specific choice of $RR$. Specifically, in all four
cases, periodic trajectories are characterized by larger values of the
different measures.

\section{Results} \label{shrimp_num} 
In the following, we study the performance of the different
recurrence-based measures for identifying shrimp structures in the
R\"ossler system. For this purpose, we take $\lambda_1$ as a
reference, since this measure per definition discriminates between
periodic ($\lambda_1=0$) and chaotic ($\lambda_1>0$) dynamics. Note
that in order to reach reliable estimates of $\lambda_1$, typically a
large amount of data is required. In contrast, recent results on RQA
and RN measures \cite{Marwan2009} suggest that these methods allow
identifying dynamical transitions in complex systems using much
shorter time series.

In order to systematically test the applicability of recurrence-based
methods for identifying shrimps, we use the same grid with $1,000
\times 1,000$ pairs of points in the $(c,a)$-parameter plane as in
Fig.~\ref{lyap_ac}. For each set of parameters, we consider time
series of $N=5,000$ data points sampled with a fixed time step of
$\Delta t= 0.2$ for estimating all RQA and RN measures. 
Initial transients {\color{red}have been} removed from the data as
described in Sec.~\ref{sec:lyap}A. We stress that measures originated
from both RQA and network theory are computed from the same recurrence
matrices $R_{i,j}$ (which have been constructed from the original
three-dimensional coordinates of the system without embedding), so
that the resulting structures in the parameter space are well
comparable. In the following, we will consistently use $l_{min}=2$ and
$RR=0.02$.

\subsection{Behavior of individual measures}
\begin{figure*}
  \centering
  \includegraphics[width=\textwidth]{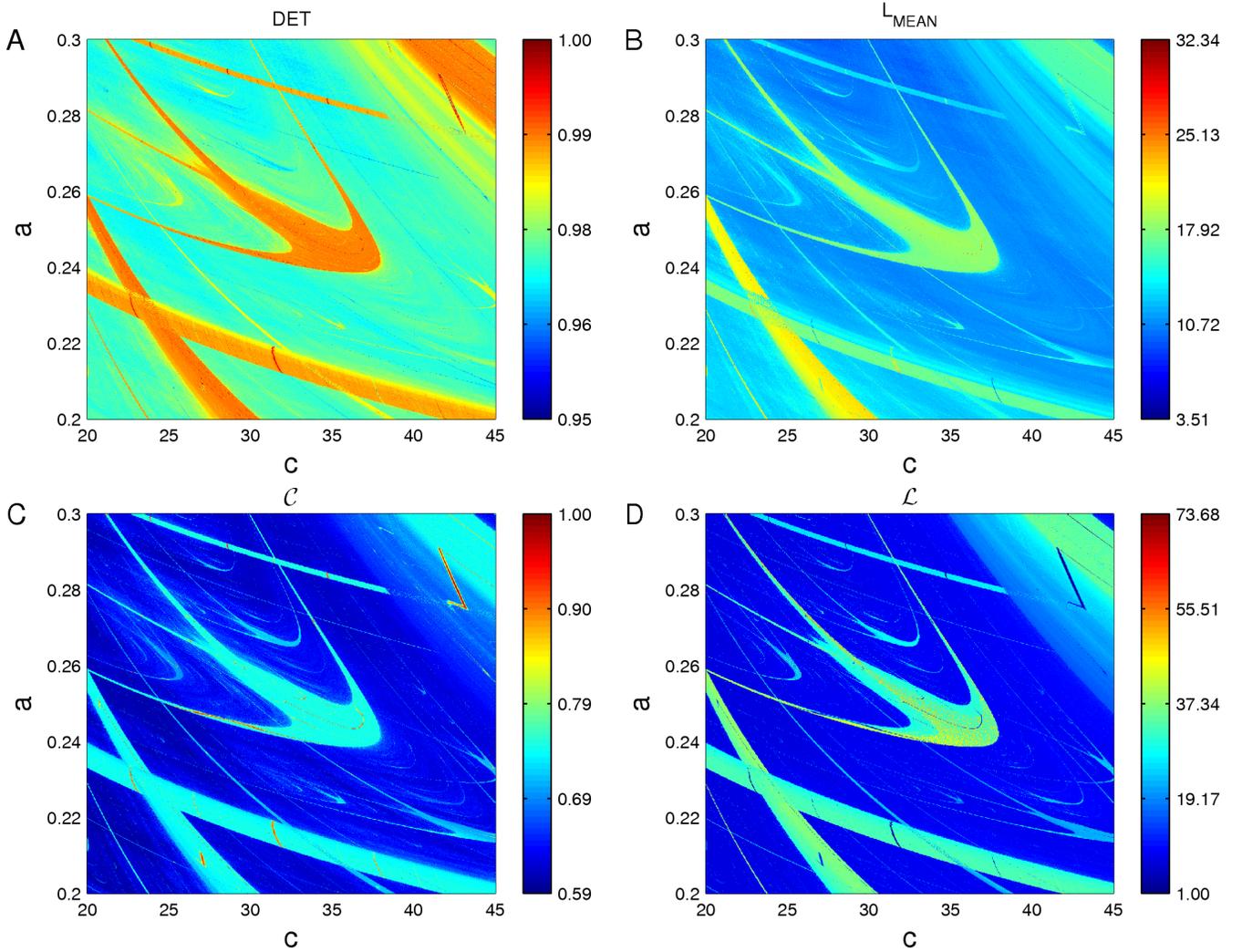}
  \caption{\small {(Color online) RQA measures $DET$ (A), $L_{MEAN}$
      (B), and network measures $\mathcal{C}$ (C) and $\mathcal{L}$
      (D) in the $(c, a)$ parameter plane of the R\"ossler system
      (\ref{ros_eqs}).}}
\label{shrimp_rqa}
\end{figure*}

The values of the four chosen measures in dependence on the parameters
$a$ and $c$ are shown in Fig.~\ref{shrimp_rqa}. We find that all four
measures are indeed able to identify periodic windows, in particular,
those associated with shrimp structures. However, the discriminatory
power of the individual parameters for distinguishing between periodic
and chaotic regions is notably different. Specifically, for the chosen
values of $l_{min}$ and $RR$ the contrast in the values of $DET$
obtained for both regimes is relatively weak (see Figs.~\ref{lmin_rqa}
and \ref{rr_rqa} and Tab.~\ref{tab_pechaotic}), whereas the other
three measures show a much larger range of values.

Unlike $DET$ and $L_{MEAN}$, $\mathcal{C}$ and $\mathcal{L}$ resolve
\textit{all} periodic regimes including band structures and even
secondary shrimps. The main remaining difference to the structures
obtained by using the maximum Lyapunov exponent (Fig.~\ref{lyap_ac})
is found in the broad periodic band in the upper right corner of the
parameter plane. In this region of parameter space, the system shows a
complex bifurcation scenario, including different routes from periodic
behavior to chaos similar to those associated with shrimp structures
in other continuous dynamical systems~\cite{Zou_ijbc_2005}. These
differences are partially related to numerical inaccuracies in the
presence of time series with a finite length and become less
pronounced if longer simulations are used. Since no analytical
solutions of the system are available for this specific region,
additional application of complementary numerical methods (e.g.,
Lyapunov exponents, return maps, Poincar\'e sections, etc.) would be
necessary for completely identifying the associated bifurcation
scenario, which is however out of the scope of this paper.

\subsection{Correlations between recurrence-based measures and
  $\lambda_1$} \label{sec:corr}
We now perform a more detailed quantitative analysis of the power of
the individual measures as discriminatory statistics between periodic
and chaotic dynamics. For this purpose, the differences between the
structures obtained using the discussed measures and those revealed by
$\lambda_1$ can be interpreted as a lack of performance. Since the
different measures are characterized by strongly differing probability
distribution functions (PDFs), we seek a statistics that allows
properly quantifying deviations between the respective structures in
the parameter space.  The first insight into this question is provided
by the correlation coefficients between $\lambda_1$ and the other
measures, which are in all cases clearly significant, but with a
negative sign (Tab.~\ref{tab_indicators}). This result supports the
findings in Sec.~\ref{sec:lyap} for the two example trajectories.

In order to study how strongly the patterns of the different measures
in the $(c,a)$-plane resemble those found for the maximum Lyapunov
exponent, we investigate the point-wise difference of the
corresponding cumulative distribution functions (CDFs), i.e.,
\begin{equation}
  \Delta P(\lambda_1,x)= P(\lambda_1)-P(x),
\label{deltap}
\end{equation}
where $P(x)$ is (for a given $(c,a)$-combination) the empirical value
of the CDF of the measure $x$ obtained from all 1,000,000 parameter
combinations. In order to simplify the notation, the arguments $(c,a)$
of $\Delta P(\lambda_1,x)$ will be suppressed. Note that the maximum
absolute value of $\Delta P$, commonly denoted $D$, corresponds to the
Kolmogorov-Smirnov statistics frequently used for testing the
homogeneity of the probability distributions of two
samples~\cite{James2006}. Since all four recurrence-based measures are
negatively correlated with $\lambda_1$, we will identify $x$ with
$1-DET$, $1-L_{MEAN}$, $1-\mathcal{C}$, and $1-\mathcal{L}$,
respectively.
\begin{table}[thb]
  \centering
  \begin{ruledtabular}
  \begin{tabular}{c|c|c}
   & $\rho$ & $\sigma^2_{\Delta P}$ \\
  \hline
  $DET$       & -0.75 & 0.21 \\
  $L_{MEAN}$  & -0.81 & 0.18 \\
  \hline
  $\mathcal{C}$ & -0.70 & 0.23 \\
  $\mathcal{L}$ & -0.66 & 0.24 \\
  \end{tabular}
  \end{ruledtabular}
  \caption{\small {Overall performance indicators obtained from a
      point-wise comparison of the values of the maximum Lyapunov
      exponent $\lambda_1$ and the different RQA and network measures:
      Spearman's $\rho$ and the standard deviation $\sigma^2$ of the CDF
      differences $\Delta P(\lambda_1,x)$. For simplicity, the arguments
      of the different characteristics have been omitted.} }
\label{tab_indicators}
\end{table}

\begin{figure*}
  \centering
  \includegraphics[width=\textwidth]{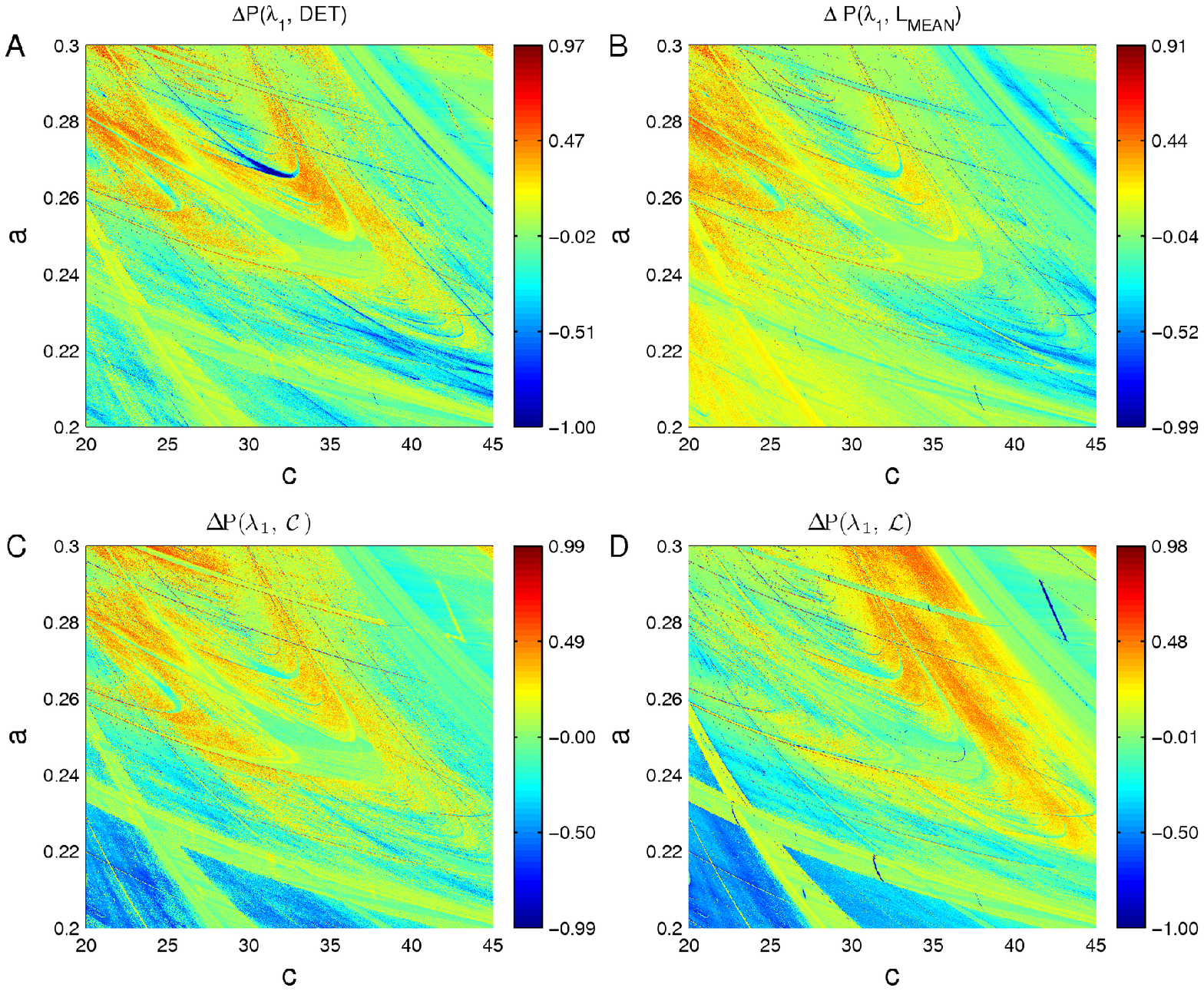} \\
  \caption{\small {(Color online) CDF differences $\Delta
      P(\lambda_1,\cdot)$ between the maximum Lyapunov exponent
      $\lambda_1$ and the RQA measures $DET$ (A), $L_{MEAN}$ (B), and
      network measures $\mathcal{C}$ (C) and $\mathcal{L}$ (D) in the
      $( c, a)$ parameter plane of the R\"ossler system
      (\ref{ros_eqs}).}}
\label{quantile_rqa}
\end{figure*}

Due to the significant correlations between the different measures
(Tab.~\ref{tab_indicators}), we expect that the CDF difference $\Delta
P(\lambda_1,x)$ is close to zero in large parts of the parameter
space. The patterns of the CDF differences are shown in
Fig.~\ref{quantile_rqa}. In all cases, values close to zero can often
be observed in periodic windows (although differences appear
especially in the secondary shrimps), whereas the shortness of the
considered time series leads to significant deviations from zero in
the chaotic regions. The standard deviations $\sigma^2_{\Delta P}$ of
the CDF field provide a rough impression of the amount of incorrectly
identified regimes (partially due to the finite length of time
series), which will be quantitatively characterized in the next
section.

\subsection{Probability of classification errors}
\begin{figure}[thb]
  \centering
  \includegraphics[scale=0.5]{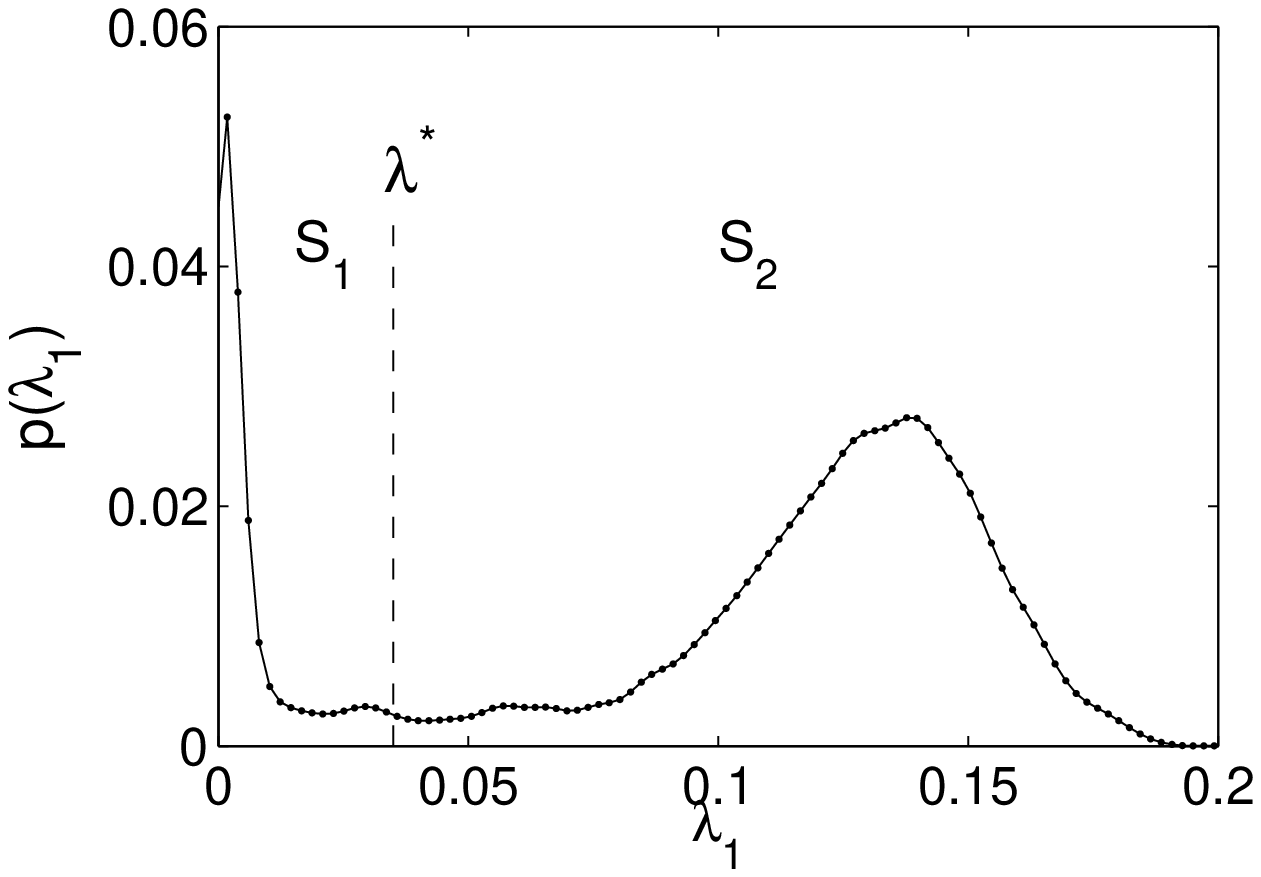}
\caption{\small {Probability distribution function of the maximum
Lyapunov exponent $\lambda_{1}$ obtained from all 1,000,000
parameter combinations in the considered $(c,a)$ plane of the
R\"ossler system (\ref{ros_eqs}).} \label{pdf_lyap} }
\end{figure}

The general treatment in Sec.~\ref{sec:corr} do not yet allow
sophisticated conclusions on which particular measure is most
appropriate for detecting shrimps in continuous systems. In order to
assess the discriminatory power of all four measures in more detail,
we subject the resulting patterns in parameter space
(Fig.~\ref{shrimp_rqa}) to further statistical analysis. For this
purpose, we explicitly make use of the fact that the $(c,a)$-parameter
plane of the R\"ossler system is composed of two sets of parameter
combinations belonging to trajectories with periodic and chaotic
dynamics, respectively. However, the corresponding group structure is
not exactly known, since there are numerous $(c,a)$-combinations which
lead to small values of $\lambda_1$ (Fig.~\ref{pdf_lyap}). These
corresponding values could hence be related to either weakly chaotic
behavior or periodic dynamics that cannot be exactly detected due to
the numerical precision. Note that no fixed points ($\lambda_1<0$) are
found in our parameter space. In order to obtain a robust
approximation of the two distinct groups of parameter combinations
leading to $\lambda_1=0$ and $\lambda_1>0$, respectively, we refer to
a (variable) critical value $\lambda^*$ (Fig.~\ref{pdf_lyap}) for
defining two disjoint sets
\begin{eqnarray*}
S_1(\lambda^*)&:=&\{(c,a)|\lambda_1(c,a) \le \lambda^*\} \\
S_2(\lambda^*)&:=&\{(c,a)|\lambda_1(c,a)>\lambda^*\}
\end{eqnarray*}
with group sizes $n_1$ and $n_2=n-n_1$ ($n=1,000,000$ in our case).

\subsubsection{$F$ and $U$ tests}
\begin{figure}[thb]
  \centering
  \includegraphics[width=0.45\textwidth]{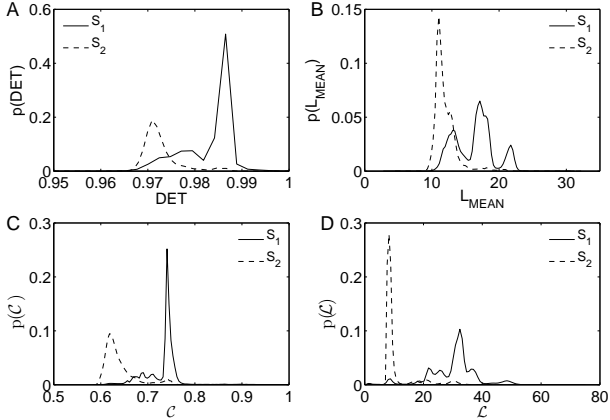}
  \caption{\small {PDFs $p(x|S_i)$ of RQA and RN measures for
      parameter combinations $(c,a)$ yielding maximum Lyapunov
      exponents $\lambda_1<\lambda^*$ (S$_1$) and
      $\lambda_1>\lambda^*$ (S$_2$) with $\lambda^*=0.01$,
      respectively.}} \label{hist}
\end{figure}

One possibility for assessing the discriminatory power of the
different recurrence-based measures is taking the respective groups
$S_1$ and $S_2$ (for different values of $\lambda^*$ in a reasonable
range, i.e., $\lambda^* \in [0,0.05]$, see Fig.~\ref{pdf_lyap}) and
statistically quantifying whether or not main statistical
characteristics of the distributions $p(x|S_i)$ of the different
measures $x$ obtained for both groups $S_1$ and $S_2$ differ
significantly. For one specific choice of $\lambda^*$, these
distributions are shown in Fig.~\ref{hist}.  Formally, this question
corresponds to a one-way analysis of variance (ANOVA)
problem~\cite{Montgomery2009}, with the factor being determined by two
classes of values of $\lambda_1$. In order to evaluate whether the
group means do significantly differ (in comparison with the respective
group variances), the $F$-test is used with the test statistics
\cite{Lomax2007}
\begin{equation}
  F = n_1 n_2 \frac{(\mu_1-\mu_2)^2}{n_1s_1^2+n_2s_2^2} \equiv t^2,
\end{equation}
where $t$ is the test statistics of a corresponding $t$-test
\cite{Welch1947}. Since we are aware that the values of $F$ (or,
alternatively, $t$) may be misleading if the underlying sample PDFs
are strongly non-Gaussian (see Fig.~\ref{hist}), our results are
complemented by those of a corresponding distribution-free test.
Specifically, we compute the value of the test statistics $U$ of the
Mann-Whitney $U$-test \cite{Conover1999,Hollander1999} against
equality of the medians of two distributions, which can be considered
as the equivalent of an $F$-test performed on the sets of rank
numbers.

\begin{figure}[thb]
  \centering
  \includegraphics[width=0.45\textwidth]{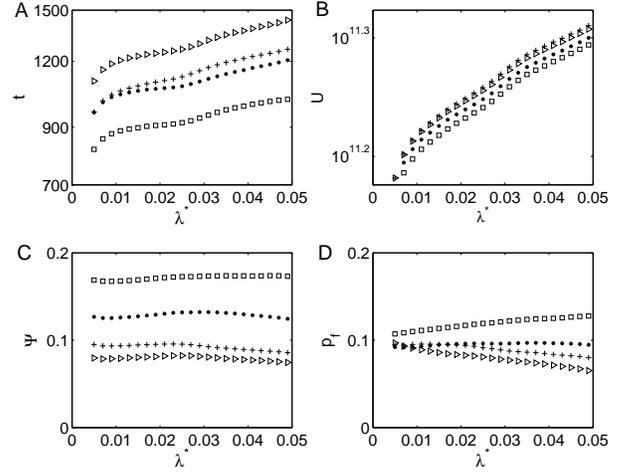}
  \caption{\small {Measures for the discriminatory skills of the
      different recurrence-based measures $DET$ ($\bullet$),
      $L_{MEAN}$ ($\Box$), $\mathcal{C}$ ($+$), and $\mathcal{L}$
      ($\triangleright$), obtained from a comparison with the results
      derived using the maximum Lyapunov exponent in dependence on the
      choice of $\lambda^*$: (A) $t$-test statistics, (B) $U$-test
      statistics, (C) overlap integral $\Psi$
      (Eq.~(\ref{eq_overlap_integral})), and (D) relative frequency of
      false detections $p_f$, using the same quantiles of $\lambda_1$
      and the respective measures.} \label{failure_p} }
\end{figure}

For both tests, we find that for almost all choices of $\lambda^*$,
$\mathcal{L}$ and $\mathcal{C}$ show the highest values of the
respective test statistics, indicating that the discriminatory power
for distinguishing between both sets is the largest for these measures
(Fig.~\ref{failure_p}A,B). Note that in all cases, the probability
values for rejecting the respective null hypotheses (i.e., equality of
group means and medians, respectively) are close to 100\% due to the
large sample size. The results obtained using both test statistics are
supported by a numerical approximation of the associated overlap
integral
\begin{equation}
  \Psi=\int_{\min(x)}^{\max(x)} dx\ p(x|S_1)\ p(x|S_2) 
\label{eq_overlap_integral}
\end{equation}
of the PDFs of the recurrence based measures $x$ for both groups
(Fig.~\ref{hist}), the values of which are shown
in~Fig.~\ref{failure_p}C.

The advantages of RN measures (at least for this particular case)
become particularly apparent when visually inspecting Fig.~\ref{hist}.
The overlap of the PDFs $p(x|S_i)$ can be seen to be substantially
smaller for the network quantifiers $\mathcal{C}$ and $\mathcal{L}$
(Fig.~\ref{hist}C,D), than for the RQA measures $DET$ and $L_{MEAN}$
(Fig.~\ref{hist}A,B).

\subsubsection{Group overlap for fixed probability quantiles}\label{quantile-grouping}
In order to further study the differences in the performance of the
considered measures, we apply another complimentary statistical test.
Specifically, we take the $\alpha$-quantile $Q_{\alpha}(\lambda_1)$ of
the distribution of $\lambda_1$ that corresponds to a given value
$\lambda^*$ (i.e., $\alpha\simeq n_1/n$) and consider a related
decomposition of the $(c,a)$ parameter plane based on the same
quantile for the recurrence-based measures, i.e.,
\begin{eqnarray*}
  S'_1(Q_{\alpha}(x))&:=&\{(c,a)|x(c,a) \le Q_{\alpha}(x)\} \\
  S'_2(Q_{\alpha}(x))&:=&\{(c,a)|x(c,a)>Q_{\alpha}(x)\}
\end{eqnarray*}
with $x\in\{1-DET,1-L_{MEAN},1-\mathcal{C},1-\mathcal{L}\}$. Since the
quantile $\alpha$ has been kept fixed here, $S'_1$ and $S'_2$ contain
the same numbers of elements ($n_1$ and $n_2$, respectively) as $S_1$
and $S_2$. Hence, we are able to quantitatively assess the coincidence
between the grouping based on $\lambda_1$ and $x$ by considering the
relative frequency $p_f$ of $(c,a)$ pairs that do \textit{not} belong
to the same group based on the two different measures.
Figure~\ref{failure_p}D demonstrates that with respect to this
criterion, the average path length $\mathcal{L}$ again shows (on
average) the lowest frequency of ``grouping errors'' in comparison to
$\lambda_1$, followed by $\mathcal{C}$, $DET$ and $L_{MEAN}$.

\subsubsection{ROC analysis}
\begin{figure}[htb]
  \centering
  \includegraphics[width=0.45\textwidth]{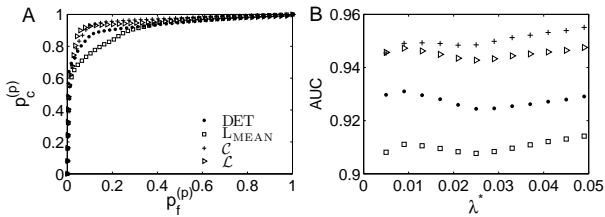}
  \caption{\small {(A) ROC curves for $\lambda^*=0.01$, and (B) area
      under the ROC curve ($AUC$) in dependence on $\lambda^*$ for all
      four measures. For $\lambda^*=0.01$, $AUC$ takes the values
      $0.9279$ ($DET$), $0.9090$ ($L_{MEAN}$), $0.9487$
      ($\mathcal{C}$), and $0.9442$ ($\mathcal{L}$), respectively.}
    \label{roc} }
\end{figure}

An even more detailed characterization of classification errors is
obtained in terms of the receiver operating characteristics
(ROC)~\cite{Fawcett2006}. In the ROC analysis, we compare the
discrimination ($S_1$, $S_2$) of the set of parameters $(c,a)$ based
on a fixed value of $\lambda^*$ with another grouping ($S'_1$, $S'_2$)
based on a variable threshold $x^*$ of the observable $x$, which now
replaces the quantile $Q_{\alpha}(x)$.  The probabilities of correct
as well as false detections of periodic behavior, $p_c^{(p)}$ and
$p_f^{(p)}$, respectively, are given as
\begin{eqnarray*}
  p_c^{(p)}(\lambda^*,x^*) &=& \left|S'_1\cap S_1\right|/\left|S_1\right| \\
  p_f^{(p)}(\lambda^*,x^*) &=& \left|S'_1\cap S_2\right|/\left|S_2\right|
\end{eqnarray*}
where $\left|S\right|$ represents the cardinality (i.e., the number of
elements) of the set $S$. Varying $x^*$ over the full range of
possible values with $\lambda^*$ simultaneously being kept fixed, we
obtain a continuous curve in the $(p_f^{(p)},p_c^{(p)})$-plane, the
ROC curve, which illustrates the trade-off between a high probability
of correct and a low probability of false detections of periodic
behavior (Fig.~\ref{roc}A). The area under this curve, $AUC$, can be
(among other statistics) used for quantitatively characterizing the
classification performance of different measures~\cite{Fawcett2006}
(Fig.~\ref{roc}B). Specifically, high values of $AUC$ correspond to a
low probability of false classifications (i.e., high specificity) and,
simultaneously, a high probability of correct classifications (i.e.,
high sensitivity). In contrast to all other kinds of statistical
tests, the ROC analysis suggests that in the specific setting studied
in this work, among the four considered measures, $\mathcal{C}$ is the
most suitable statistics for discriminating between periodic and
chaotic behavior of the R\"ossler system, followed by $\mathcal{L}$,
$DET$, and $L_{MEAN}$.

\begin{figure}[htb]
  \centering
  \includegraphics[width=0.45\textwidth]{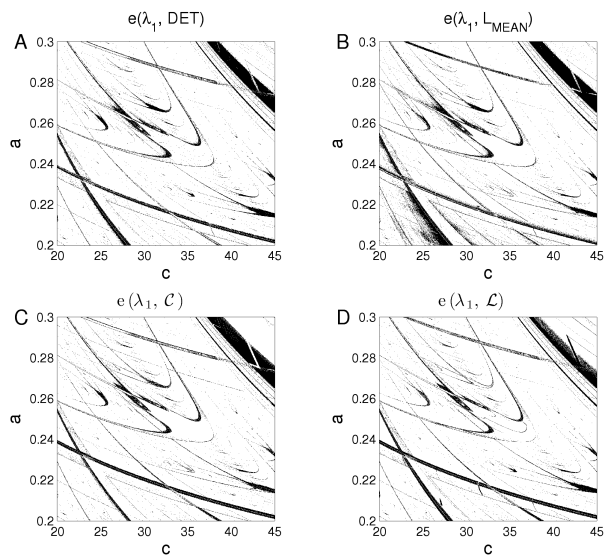}
  \caption{\small {Discrimination errors (black dots) for the
      quantile-based groupings $S_1$, $S_2$, $S'_1$, and $S'_2$ for
      $\lambda^*=0.01$ (see Sect.~\ref{quantile-grouping}) for (A)
      $DET$ ($p_f=0.0923$), (B) $L_{MEAN}$ ($0.1106$), (C)
      $\mathcal{C}$ ($0.0954$), and (D) $\mathcal{L}$ ($0.0899$).}
\label{errorpattern} }
\end{figure}

One has to note that the overall classification error
\begin{equation}
  \begin{split}
    p_f&=p_f^{(p)}(\lambda^*,Q_{\alpha}(x))+p_f^{(c)}(\lambda^*,Q_{\alpha}(x)) \\
    &=p_f^{(p)}(\lambda^*,Q_{\alpha}(x))+\left(1-p_c^{(p)}(\lambda^*,Q_{\alpha}(x))\right)
 \end{split}
\end{equation}
(with $p_f^{(c)}(\lambda^*,Q_{\alpha}(x))=|S'_2\cap S_1|/|S_1|$ being
the relative frequency of false detections of chaotic dynamics) still
reaches values of around $10\%$ even for the best performing measures
(Fig.~\ref{failure_p}D). The main reason for this performance failure
is the different sensitivity of $\lambda_1$ and the recurrence-based
measures close to the bifurcation lines (see Fig.~\ref{errorpattern}).
Apart from this, all four measures allow recovering the overall
structures in parameter space comparably well as $\lambda_1$. The
obvious differences in the transition regions could be related to the
use of short time series in regions with a complex bifurcation
scenario, which are probably affected by problematic features such as
longer transients before the attractor is reached (see
Sec.~\ref{sec:net}), different bifurcation scenarios characterizing
the transition between periodic and chaotic dynamics across different
boundaries of the same shrimp, intermittency, or different stiffness
properties of the considered
trajectories~\cite{Gallas1,Zou_ijbc_2005}.

\section{Conclusions} \label{sec:conclusion} We have proposed to use
nonlinear recurrence-based characteristics of time series for
exploring the parameter space of complex systems. This is of special
importance when dealing with experimental data. Specifically, for
distinguishing periodic and chaotic dynamics, in the recent literature
estimates of the maximum Lyapunov exponent $\lambda_1$ from the
corresponding ODEs have most often been used, which allow resolving
the borders of shrimps in a satisfactory way (see Fig.~\ref{lyap_ac}).
This specific approach works well if the associated variational
equations are provided explicitly. However, if these equations are not
available (as in the case of experimental data), the
\textit{numerical} estimation of $\lambda_1$ is typically much more
challenging, especially when dealing with short time
series~\cite{Kantz97}. The recurrence-based measures used in this work
have the advantage that they can be properly estimated from rather
short time series, so that they could prove advantageous also in
situations where the available amount of data is not sufficient for
obtaining reliable estimates of Lyapunov exponents. Using
recurrence-based methods instead of Lyapunov exponents therefore has
great potentials for the automatized discrimination between different
types of dynamics in many applications. Specifically, the alternative
discriminatory statistics introduced in this work could become very
helpful not only in evaluating, but also already in designing both
experimental and numerical studies, since the requirements concerning
the necessary amount of data can be much easier matched. As a
perspective, we therefore expect that systematic application of the
proposed methods will open new fields of applications, as it has
already been the case with the introduction of
RQA~\cite{Marwan_report_2007}. However, there are still situations
where the consideration of Lyapunov exponents is clearly superior for
detecting periodic windows in comparison to the recurrence-based
approach, in particular when the governing equations are explicitly
known or the available observations of the considered systems are very
long. It will remain a subject of future studies to investigate in
more detail under which specific conditions which of the two
approaches is favorable.

Another traditional idea to distinguish periodic behavior from chaotic
dynamics is to use a properly defined surface in phase space for a
Poincar\'e section, since a periodic trajectory has only a finite
number of intersection points with this surface, while a chaotic one
renders an infinite number of crossings. Note, however, that
constructing a proper Poincar\'e section for a given set of
ODEs~\cite{HenonP_phyD_1982} is much easier than it is the case when
working with time series, where one most often relies on some
interpolation techniques to find crossings on the section, which
obviously introduces noise-like effects~\cite{Kantz97}. Furthermore,
there are no universal criteria for choosing Poincar\'e sections when
scanning a large parameter space (e.g., Fig.~\ref{lyap_ac}) since the
shape and the orientation of the attractors typically vary for
different parameter values.

In this work, we have been particularly interested in the problem of
detecting specific periodic windows in parameter space, so-called
shrimps, which are characterized by a rich bifurcation scenario and
self-similarity. Specifically, we have addressed the problem of
numerically detecting shrimps in systems of ODEs, while the related
question of the associated bifurcation scenarios remains a subject for
future investigations. We expect the recurrence-based methods proposed
in this work to be particularly useful for this problem, especially
concerning the properties of secondary shrimps and the quantitative
analysis of possible transient behavior.

For properly identifying shrimps in parameter space based on
recurrence plots, both measures from RQA and complex network theory
have been applied. For the $(c,a)$ parameter plane of the R\"ossler
system, especially the recently proposed application of network
measures to the recurrence matrix of complex
systems~\cite{Marwan2009,Donner2009} yields results coinciding well
with those obtained using $\lambda_1$. The used RQA measures in this
paper need two parameters recurrence rate $RR$ and minimum line length
$l_{min}$, RN measures depend exclusively on $RR$.  We have to
emphasize that this conceptual advantage comes at the cost of higher
computational demands, especially when considering the average path
length $\mathcal{L}$ of the RNs. However, the clustering coefficient
$\mathcal{C}$, which has been found to perform best in our example
when considering the most sophisticated statistical evaluation (ROC
analysis), can be calculated at significantly lower costs.  Generally,
$\mathcal{C}$ seems to be very well suited for distinguishing periodic
from chaotic dynamics, even for time series sampled at a very high
rate, such as the examples presented in this work. However, for high
sampling rates, $DET$ and $L_{MEAN}$ can have a reduced discriminatory
performance since the typical line structures become too
similar~\cite{Marwan2011}. We additionally note that both network
measures have shown a slightly better discriminatory power for
secondary shrimps than the two RQA measures $DET$ and $L_{MEAN}$, at
least for the specific setting used in this study.  As we have not
explicitly shown here, even larger performance errors can be observed
for other line-based RQA measures. A more detailed investigation of
possible impacts of other choices of $l_{min}$ and $RR$ has not yet
been systematically performed, but will be subject of future studies.
As a preliminary result, repeating all presented calculations with a
different value of $RR=0.01$ supports all results discussed in detail
for $RR=0.02$ in this paper.

The two specific network-theoretic quantities considered here have
already been shown to be able to detect dynamical transitions in both
model systems and real-world time series~\cite{Marwan2009}.  We note,
however, that since both measures are invariant under permutations of
the time coordinate~\cite{Donner2009}, they exclusively capture the
geometry of states associated to a specific trajectory in phase space.
This is a distinctive difference to most existing methods of time
series analysis, which are related to the study of temporal
correlations.  We hypothesize that the good performance of network
measures in detecting shrimps could be related to this fact, since
network measures take only information about spatial correlations
(i.e., neighborhood relationships in phase space) into account.

While a rigorous theory interrelating the phase space properties
captured by RN measures to traditional dynamical invariants is still
missing, the algorithm for estimating the correlation entropy $K_{2}$
from RPs (which has also been successfully applied for detecting
shrimps in continuous systems~\cite{MTthesis}) has been justified
theoretically~\cite{fluid}. Since the estimation of $K_{2}$, however,
involves several steps~\cite{MTthesis} with some subjective issues (in
particular, the choice of the scaling region for convergence), its
computational demands are significantly higher than those of the
complex network as well as RQA measures applied in this work.
Particularly, a large range of values of $RR$ ($1 \% \sim 99 \%$) is
often considered to yield a well defined plateau of $K_{2}(RR)$, while
the method proposed in this paper works with just one chosen value of
$RR$, dramatically reducing computational costs. Systematically
calculating RN statistics as well as RQA measures can be easily
automatized using a fixed recurrence rate $RR$, ideally with
$RR\lesssim 0.05$ \cite{Schinkel2008,Donner2009b}. Hence, we conclude
that our approach allows a systematic numerical discrimination between
periodic and chaotic dynamics of a continuous system, which is more
practicable than other possible techniques especially when
systematically studying higher-dimensional parameter spaces.

In relation with the problem of discriminating between two
qualitatively different types of dynamics in a binary way, which has
been discussed in this work, Lyapunov exponents have also found wide
use in quantitatively characterizing the ``chaoticity'' of complex
dynamics. Following the results from Tab.~\ref{tab_indicators}, we
emphasize that the recurrence-based characteristics considered in this
work may be used for similar purposes. In contrast to the
discriminatory power, our corresponding initial results suggest that
the applied RQA measures may be somewhat better suited for this
purpose than the network-based concepts. However, further detailed
statistical analysis is necessary in order to provide further evidence
that this result holds in general. A detailed treatment of this
question therefore remains subject of future studies.

\acknowledgments{This work has been financially supported by the
  German Research Foundation (DFG project no.  He 2789/8-2), the Max
  Planck Society, the Federal Ministry for Education and Research
  (BMBF) via the Potsdam Research Cluster for Georisk Analysis,
  Environmental Change and Sustainability (PROGRESS), and the Leibniz
  association (project ECONS). All complex network measures have been
  calculated using the software package \texttt{igraph}
  \cite{Csardi2006}. We thank K. Kramer for help with the IBM
  iDataPlex Cluster at the Potsdam Institute for Climate Impact
  Research, on which all calculations were performed. }


\end{document}